\documentclass[twocolumn]{aastex631}

\DeclareUnicodeCharacter{2212}{-}
\usepackage{soul}

\shorttitle{TOI-251 in Pheonix-Grus}
\shortauthors{Sun et al.}

\graphicspath{{./}{figures/}}
	
\begin{document}
		
		\title{A search for stellar siblings of the $\sim$ 200 Myr TOI-251b planetary system}
		
		\correspondingauthor{Qinghui Sun}
		\email{qingsun@mail.tsinghua.edu.cn}
		
		\author[0000-0003-3281-6461]{Qinghui Sun}
		
		\author[0000-0002-6937-9034]{Sharon Xuesong Wang}
		\affiliation{Department of Astronomy, Tsinghua University, Beijing, 100084, China}
		
		\author[0000-0003-3654-1602]{Andrew W. Mann}
		\affiliation{Department of Physics and Astronomy, The University of North Carolina at Chapel Hill, Chapel Hill, NC 27599, USA}
		
		\author[0000-0003-2053-0749]{Benjamin M. Tofflemire}
		\altaffiliation{51 Pegasi b Fellow}		
		\author[0000-0001-9811-568X]{Adam L. Kraus}
		\affiliation{Department of Astronomy, The University of Texas at Austin, Austin, TX 78712, USA}
		
		\author[0000-0002-4503-9705]{Tianjun Gan}
		\affiliation{Department of Astronomy, Tsinghua University, Beijing, 100084, China}
		
		\author[0000-0002-4503-9705]{Madyson G. Barber}
		\affiliation{Department of Physics and Astronomy, The University of North Carolina at Chapel Hill, Chapel Hill, NC 27599, USA}
		
		\begin{abstract}
			
			Young planets ($<$ 1 Gyr) are helpful for studying the physical processes occurring at the early stage of planet evolution. TOI-251b is a recently discovered sub-Neptune orbiting a young G dwarf, which has an imprecise age estimation of 40-320 Myr. We select TOI-251 sibling candidates based on kinematics and spatial proximity to TOI-251, and further use the color-magnitude diagram (CMD) to refine the list and to compare to multiple open clusters. We report stellar rotational period for 321 sibling candidates in a 50 pc radius around TOI-251 by analyzing their stellar light curves, and find a color -- rotational period sequence that lie in between the Group X (300 Myr) and Pleiades (120 Myr) members, suggesting an age $\sim$ 200 Myr. A quantitative age analysis by using gyrochronology relations give 204 $\pm$ 45 Myr, consistent with the average Li-age of selected siblings (238 $\pm$ 38 Myr) and the Gaia variability age (193$^{102}_{-54}$ Myr). The detection fraction of comoving candidates that have short rotational period is 68.1\%, much higher than the typical value in the field (14\% - 16\% from Kepler). The overdensity of young stars and consistency in age of stellar siblings suggest a potential young association candidate in the Pheonix-Grus constellation. Though TOI-251 b has a radius larger than most of its field-age counterparts, we are uncertain whether TOI-251 is inflated due to a lack of knowledge on the planet's mass.
			
		\end{abstract}
		
		\section{Introduction} \label{sec:intro}
		
		Planets evolve quickly in the first hundred million years after they form (\citealt{2013ApJ...775..105O}, \citealt{2013ApJ...776....2L}) due to major physical processes in the formation process such as planet migration (\citealt{2017AJ....154..106N}) and photoevaporation (\citealt{2008MNRAS.384..663R}, \citealt{2018MNRAS.479.5012O}). Young planets ($<$ 1 Gyr) of different ages will help to understand the timing of early planet evolution, and very young planets ($<$ 100 Myr) can even tell us the physical processes right after they form (\citealt{2020AJ....160...33R}, \citealt{2016AJ....152...61M}, \citealt{2019ApJ...885L..12D}, \citealt{2016Natur.534..658D}, \citealt{2016ApJ...826..206J}).
		
		The number of newly discovered young transiting planets has grown significantly in recent years. A majority of the exoplanets discovered so far are around field stars (from the NASA Exoplanet Archive\footnote{\url{https://exoplanetarchive.ipac.caltech.edu/}}), while in the meantime an increasing number of planets are found in open clusters (e.g. \citealt{2014ApJ...787...27Q}, \citealt{2012ApJ...756L..33Q}, \citealt{2016AJ....152..223O}, \citealt{2020AJ....160..239B}) and in young associations, e.g., the ones reported by the TESS Hunt for Young and Maturing Exoplanets (THYME) group \citep{2019ApJ...880L..17N, 2020AJ....160...33R, 2020AJ....160..179M, 2021AJ....161...65N, 2021AJ....161..171T, 2022AJ....163..156M}. The rapid growth of new discoveries of young planets is primarily driven by NASA's Kepler/K2 mission (\citealt{2010Sci...327..977B}, \citealt{2014PASP..126..398H}) and the TESS survey (\citealt{2015JATIS...1a4003R}), which provided a long-baseline, high-precision photometric coverage on a large sample of young stars for the first time and enabled discoveries of transiting planets around them. The TESS survey has a much larger sky coverage than Kepler/K2 and in the meantime goes to fainter young stars thanks to its Full Frame Images (FFIs). For example, the TESS Quick-Look Pipeline (QLP) faint star search using the FFIs (\citealt{2022ApJS..259...33K}) include stars as faint as $T$ = 13.5 mag (in TESS magnitude). The growing number of planets in stellar associations is also driven by the advancements in the open cluster and young association catalogs (\citealt{2018AA...618A..93C}, \citealt{2022arXiv220305177H}, \citealt{2018ApJ...856...23G}, \citealt{2021ApJ...917...23K}).
		
		Open clusters and young associations are coeval populations, whose members have similar movements, chemical compositions and age, making them superior to field stars for population studies (\citealt{2021PhDT........18S}). It is easier to track their evolutionary history and derive well-constrained stellar parameters, including age, which can be important for planetary population studies. Despite the growing number of planets found in coeval populations, most of the published young planets so far are around field stars rather than association member stars. Based on the idea that stars form in clustering environments (\citealt{2003ARAA..41...57L}), more young planets are expected to be in open clusters or young associations. The number of young planets discovered may be constrained by the magnitude limit of planet surveys, as well as the incompleteness of stellar association catalogs. Open clusters have typical ages between a few Myr to a few Gyr, young associations are usually younger with typical ages between 1--800 Myr (\citealt{2018ApJ...856...23G}), so it is promising to find very young planet with well-determined age in them.
		
		The standard searches of planets in young associations may miss some of the candidates. \citet{2022RAA....22g5008S} recently used the BANYAN $\Sigma$ Bayesian algorithm to search for planets in 27 young association and has successfully matched several TOIs with young associations. However, the number is still low and their affiliations with the young associations are occasionally wrong. In this work we present a different approach to identify stellar siblings and young association candidates where a planet resides, whcih in turn helps to constrain its age. This method has successfully identified the 250 Myr MELANGE-1 (\citealt{2021AJ....161..171T}), MELANGE-2 (\citealt{2022AJ....164..115N}), and the Pleiades-aged MELANGE-3 association (candidate) (\citealt{2022AJ....164...88B}). The increasing number of planets found in stellar associations would help to answer whether the planet occurrence rate in stellar associations shows any difference from that in the field stars, which is essential for understanding the early planet formation and evolution conditions (\citealt{2021AJ....162...46D}). 
		
		\citet{2021AJ....161....2Z} discovered TOI-251b -- a sub-Neptune orbiting around a G-type star. The planetary system was not knowingly belong to any known coeval stellar population, leading to an imprecise age estimation of 40-320 Myr. If the age of the planetary system lies closer to the lower boundary of this range, then the system would be especially helpful for understanding physical processes occurring inside the planets right after they form. A more precise age measurement is thus desirable. In this work, we search TOI-251's stellar neighbors based on kinematics and spatial proximity (section \ref{sec:sibling}), and compare to the color-magnitude diagram of multiple open clusters (section \ref{sec:CMD}). We then derive stellar rotational period ($P_{ROT}$) for all sibling candidates, estimate their gyrochronology age (section \ref{sec:rotp}), derive the Li-age of selected stellar siblings (section \ref{sec:li}), and compute the Gaia variability age (section \ref{sec:gaia_age}). The siblings of TOI-251 suggest a homogenous age $\sim$ 200 Myr, and a young association candidate in the Pheonix-Grus constellation (section \ref{sec:new_association}).
		
		\section{TOI-251 and its neighborhood}  \label{sec:251}
		
		\subsection{Selection of TOI-251 sibling candidates}  \label{sec:sibling}
		
		\citet{2021AJ....161....2Z} find a sub-Neptune orbiting around the G type star TOI-251 (also TIC 224225541). We use stellar parameters directly from \citet{2021AJ....161....2Z} in our work, some of the relevant parameters are also listed in Table \ref{tab:251friends} and \ref{tab:rotp}. TOI-251 has an effective temperature of $T_{\rm eff}$ =  5875$^{+100}_{−190}$ K and a metallicity of [Fe/H] = −0.106$^{+0.079}_{−0.070}$ dex. The sub-Neptune TOI-251b has an orbital period of 4.937770$^{+0.000028}_{−0.000029 }$ days and a semi-major axis of 0.05741$^{+0.00023}_{−0.00017}$ AU. The radius and mass of the planet are 2.74$^{+0.18}_{-0.18}$ R$_E$, and $<$ 1.0 M$_{Jup}$, respectively.
		
		TOI-251 exhibits significant photometric variability, significant chromospheric activity, and significant Lithium absorption feature, so it is very likely to be of a young age. A more precise age estimation is possible if the system is in a young association. Given its young age, we search TOI-251 neighborhood for potentially new young associations.
		
		We use the \texttt{Comove}\footnote{https://github.com/adamkraus/Comove} python package to search for TOI-251 neighborhood based on kinematics and space proximity (potential siblings, or comoving candidates). To begin with, the program will search candidates within a given radius around TOI-251 by using the Gaia Data Release 3 (DR3, \citealt{2022arXiv220605989B, https://doi.org/10.26131/irsa544}) catalog, and retrieve the candidates' stellar parameters (e.g. magnitude, color, parallax, radial velocity, proper motion, etc.). It then uses the Gaia proper motion to compute the UVW space velocity of the target star (e.g., \citealt{1987AJ.....93..864J}), and projects the target star’s UVW velocity to the tangential and radial velocity assuming all the candidates are comoving at the same UVW. For all the initial candidates, the program computes the tangential velocity offset (V$_{off}$) and accept those within a given V$_{off}$ threshold. Finally, \texttt{Comove} queries the Two Micron All Sky Survey (2MASS, \citealt{2006AJ....131.1163S, https://doi.org/10.26131/irsa2}), the Galaxy Evolution Explorer (GALEX, \citealt{2017ApJS..230...24B, https://doi.org/10.26131/irsa166}), the ROSAT all-sky survey (\citealt{2016AA...588A.103B}), and the Wide-field Infrared Survey Explorer (WISE, \citealt{2012yCat.2311....0C, https://doi.org/10.26131/irsa1}) catalog for additional information such as the $W1-W3$ color (from WISE) and the NUV/NIR flux ratio (F$_{NUV} $/F$_J$, from GALEX). 
		
\begin{figure*}[ht!]
			\centering
			\includegraphics[width=0.8\textwidth]{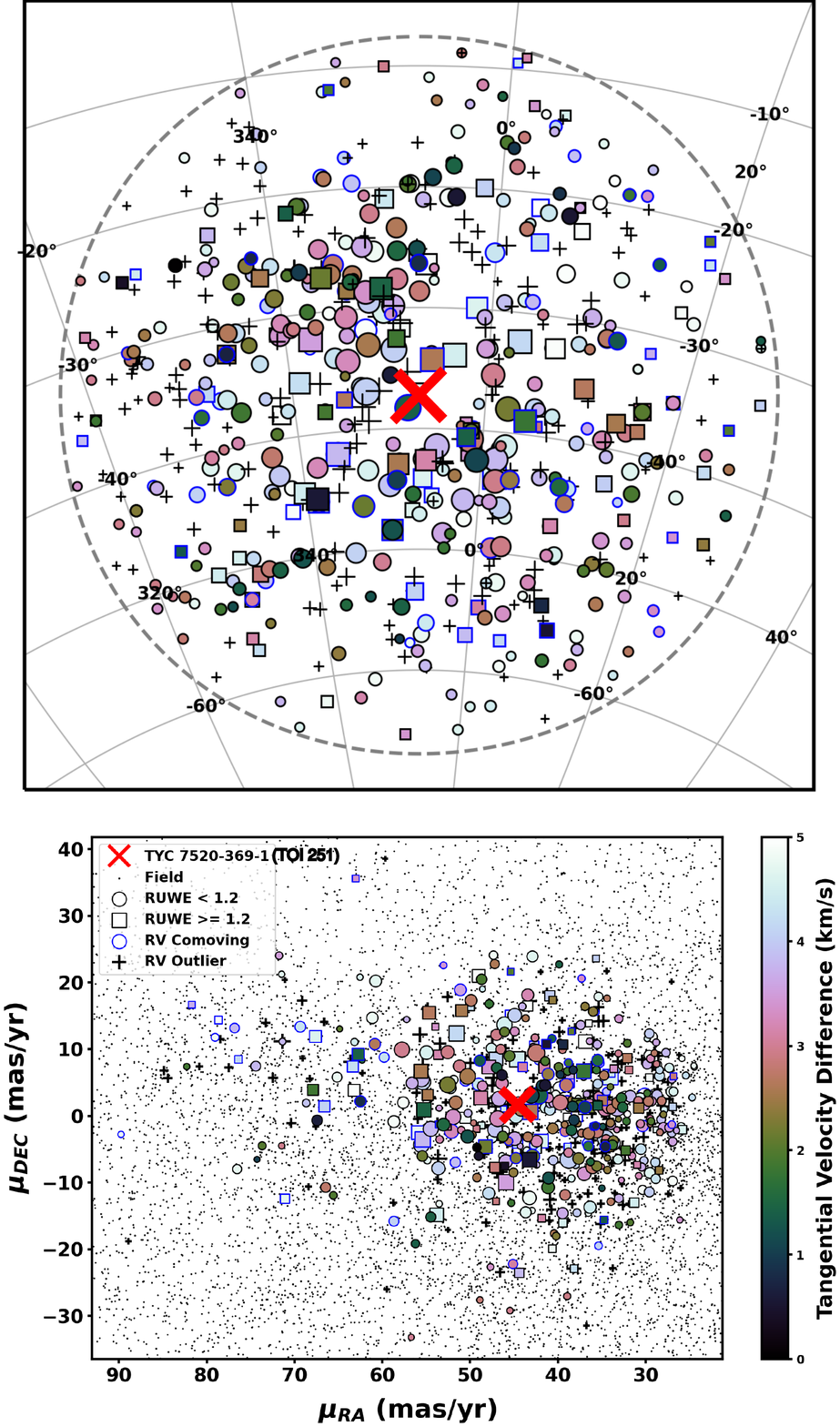} 
			\caption{TOI-251 neighbors from \texttt{Comove}. The top panel shows the neighbors' 3D spatial distribution, the bottom panel shows the neighbors' proper motion distributions in RA ($\mu_{RA}$) and DEC ($\mu_{DEC}$). TOI-251 is in a red cross, all the neighboring candidates are located within a spherical volume of R $<$ 50 pc. Symbol sizes are proportional to the 3D distance from TOI-251, and colored by their tangential velocity offset ($V_{off}$). Objects marked in circles have RUWE $<$ 1.2 and are likely single stars, and squares are used for likely binary stars with RUWE $>$ 1.2. If an object is shown in a + symbol, its radial velocity from Gaia DR3 is discrepant by $>$ 5 km s$^{-1}$, and hence it is assumed not to be associated. Also in the right panel, the small dots show objects that fall within the spatial search volume but do not have a tangential velocity offset within $<$ 5 km/s of TOI-251.}
			\label{fig:spatial}
		\end{figure*}

		We adopt coordinates and radial velocity of TOI-251 from its discovery paper (also shown in the first row of Table \ref{tab:251friends}), set a maximum spatial radius of 50 pc and V$_{off}$ $<$ 5 km s$^{-1}$ as the input parameters for \texttt{Comove}. As a result, the program returns to us 626 candidates around TOI-251. Figure \ref{fig:spatial} provides an overview of the search result directly from \texttt{Comove}. The top panel shows the spatial distributions of the initial candidates. The symbol sizes are proportional to their 3D distances to TOI-251, colored by their tangential velocity offset ($V_{off}$). The Gaia renormalized unit weight error (RUWE), or the reduced $\chi^2$ of the single-source astrometric fit, can serve as an indicator of the source's multiplicity (\citealt{2020MNRAS.496.1922B}). Stars that are most likely single (RUWE $<$ 1.2) are shown in circles, while potential binary stars (RUWE $>$ 1.2) are shown in squares. Some (but not all) candidates have radial velocities (RV) reported by Gaia DR3, among them we mark objects that have RV offset $>$ 5 km s$^{-1}$ (difference from the RV of TOI-251) in plus signs (``+'', unlikely to be associated). The bottom panel shows the Gaia proper motion distributions of initial candidates, the same symbols are used as in the top panel. Note that we do not use proper motion as a cut here and the bottom plot is just included here for illustration.
		
\begin{figure*}[ht!]
			\centering
			\includegraphics[width=1.0\textwidth]{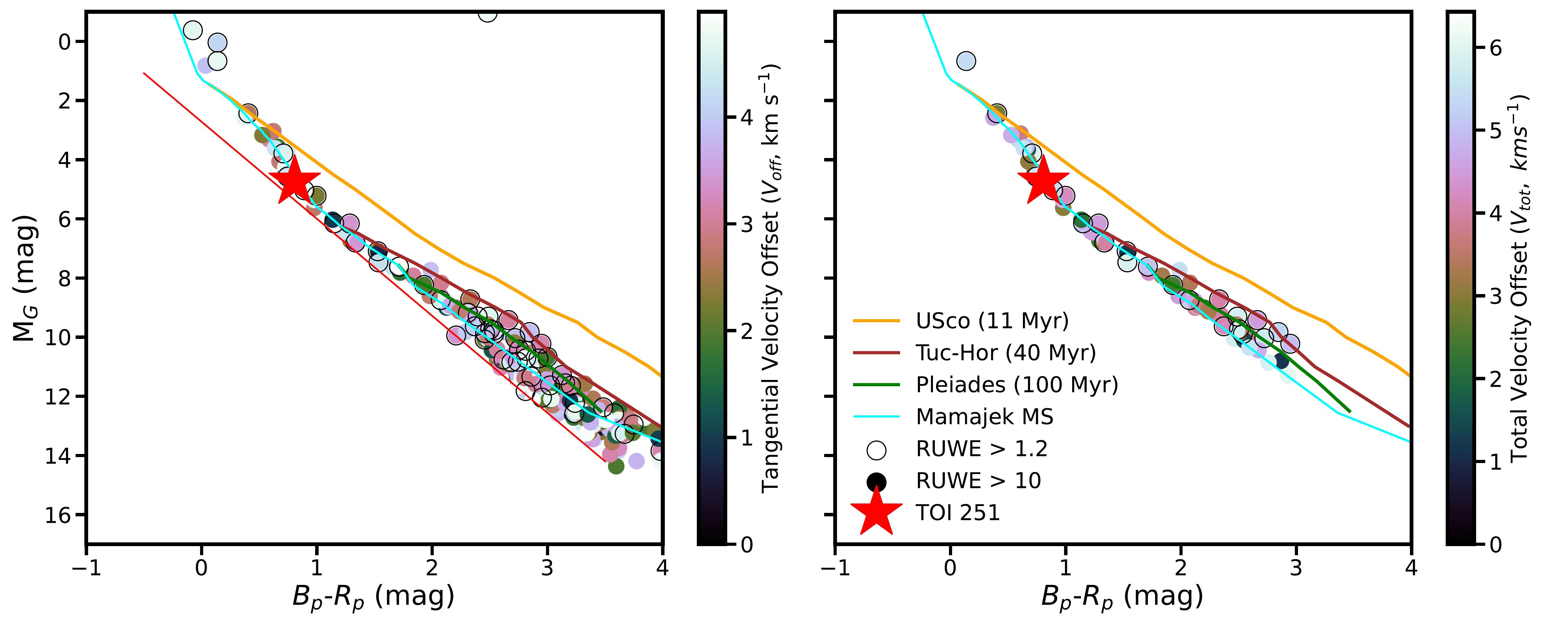} 
			\caption{The left panel is a color–magnitude diagram of TOI-251 neighbors (within 50 pc, excluding stars with RV offset $>$ 5 km s$^{-1}$), which are colored by their $V_{off}$s. Objects that fall off the main-sequence (below the red line) are excluded for further analysis. The right panel show only candidates that have reported Gaia DR3 RVs, colored by their total velocity offsets (computed from RV and $V_{off}$). Stars with excessively large RUWE ($>$ 10, filled black circles) are excluded for further analysis. Candidates with RUWE $>$ 1.2 are likely unresolved binaries, we encircle them to distinguish from single young stars (RUWE $<$ 1.2).}
			\label{fig:CMD1}
		\end{figure*}

		We then take a few steps to clean the initial candidate list and select comoving candidates. We first exclude objects that have RV offset $>$ 5 km s$^{-1}$, after this cut, we have 406 objects remaining (including TOI-251). Candidates that have no RV information from Gaia DR3 are still included in Table \ref{tab:251friends} because they have the chance to be associated, and further RV (and/or other) information are needed to confirm or reject this. The left panel of Figure \ref{fig:CMD1} is a color-magnitude diagram (CMD) showing the 406 objects, using the Gaia absolute magnitude ($M_G$) and $B_p-R_P$ color. $M_G$ is computed based on the Gaia $G$ magnitude and parallax. We first exclude all 77 stars below the red line, which include the apparent non-members that fall off the main-sequence (especially toward the fainter end), and candidates on the white dwarf sequence as they are more likely contaminants in the field (none of them have RV). The remaining ``good" candidates all lie along (or close to) the main-sequence, that can be directly compared to other young clusters to determine their age. We further refine our sample by excluding the eight candidates with excessively large RUWE ($>$ 10, shown as filled black circles). Candidates with RUWE $>$ 1.2 are likely unresolved binaries, so we distinguish them from single young stars by encircling them. The objects are colored by their $V_{off}$s. The right panel of Figure \ref{fig:CMD1} shows only stars that have reported Gaia DR3 RV, and the same symbols are used as in the left panel except for that the objects are colored by their total velocity offsets ($V_{tot}$, computed from RV and $V_{off}$). The candidate stars show a clearly defined main-sequence that falls in between the Tuc-Hor (\citealt{2014AJ....147..146K}) young association and the Mamajek sequence (typical main sequence for the Solar Neighborhood, \citealt{2013ApJS..208....9P}). 
		
		From the discussions above, we exclude the 77 photometric nonmembers, and eight stars that have very large RUWE ($>$ 10) from the original candidate list. There are 321 objects remaining and we call them TOI-251 sibling candidates. Table \ref{tab:251friends} summarizes the basic parameters for all the sibling.candidates. Column 1 is the Gaia DR3 ID, Column 2 is the TESS Input Catalog ID (TIC, \citealt{2018AJ....156..102S}), Columns 3-20 are RA and DEC, $G$ magnitude,$R_p$, $B_p-R_p$, Bp/Rp flux excess, tangential velocity offset (km s$^{-1}$), spatial separation (degree), 3D distance (pc), predicted RV (km s$^{-1}$), observed RV (km s$^{-1}$), error in RV (km s$^{-1}$), total velocity offset ($V_{tot}$), parallax (mas), spectral type, F$_{NUV} $/F$_J$, $W1-W3$, and RUWE, respectively. Columns 21-23 are rotational period ($P_{ROT}$), error in $P_{ROT}$, and notes on the light curves. Among the 321 objects, 263 are likely single stars (RUWE $<$ 1.2 , including TOI-251), and 58 are likely binaries (RUWE $> 1.2$). Only 73 candidates (22.7\%) have Gaia DR3 RV, we call them sibling (or RV comoving) candidates and use them for age analysis in later sections. Other candidates that do not have Gaia DR3 RV are all marked as uncertain siblings (``U") in column 24 of Table \ref{tab:251friends}. We finally confirm 11 stellar siblings of TOI-251, and show the entire rows for them in Table \ref{tab:251friends}. Other uncertain siblings and non-siblings are not shown explicitly here, but included in the entire table that is available online.
		
		\begin{splitdeluxetable*}{ccccccccccBcccccccccccccc}
			\tablecaption{Parameters for TOI-251 sibling candidates} \label{tab:251friends}
			\tiny
			\tablehead{
				\colhead{Gaia DR3} & \colhead{TIC} & \colhead{RA} & \colhead{DEC} & \colhead{$G$}	& \colhead{$R_p$}	& \colhead{\scriptsize{$B_p-R_p^2$}}	& \colhead{$B_p-R_p$} & \colhead{$V_{off}$} & \colhead{Sep} & \colhead{3D} & \colhead{Vr(pred)} & \colhead{Vr(obs)} & \colhead{Vrerr} & \colhead{$V_{tot}$} & \colhead{plx} &  \colhead{SpT} &  \colhead{F$_{NUV} $/F$_J$} &  \colhead{\scriptsize{$W1-W3$}} &  \colhead{\scriptsize RUWE} &  \colhead{$P_{ROT}$} &  \colhead{$\sigma(P_{ROT})$} &\colhead{\scriptsize Notes} &  \colhead{\scriptsize{sibling$^3$}}  \\
				\colhead{ID} & \colhead{ID} & \colhead{(J2000)} & \colhead{(J2000)} & \colhead{mag}	& \colhead{mag}	& \colhead{\scriptsize{mag}}	& \colhead{excess} & \colhead{$km s^{-1}$} & \colhead{deg} & \colhead{pc}	& \colhead{$km s^{-1}$}	& \colhead{$km s^{-1}$}	& \colhead{$km s^{-1}$} & \colhead{$km s^{-1}$} & \colhead{mas} & \colhead{} & \colhead{} & \colhead{mag} & \colhead{} & days & days & \colhead{} & \colhead{} 
			}
			\decimalcolnumbers
			\startdata
			6539037542941988736 (TOI-251) & 224225541 & 353.0620734 & -37.2558627 &  9.761 &  9.291 & 0.807 & 1.182 & 0.000 &  0.0000 &  0.0000 & -2.33 & -1.52 &  0.36 & 0.00 &  9.901 & G1.1 &      nan &  0.000 & 0.959 & 3.84 & 0.51  & --  & Y \\
			6550497688214589440 & 175490413 & 351.8561114 & -38.2739063 &  9.638 &  9.206 & 0.730 & 1.180 & 1.425 &  1.3947 & 10.6152 & -2.70 &  1.16 &  0.32 & 3.04 &  8.985 & F9.2 &      nan &  0.000 & 0.948 & 3.29 & 0.47  & --  & Y \\
			6556983879105150848 & 270585410 & 347.8721477 & -31.7040405 & 11.463 & 10.830 & 1.136 & 1.223 & 4.926 &  7.0062 & 19.7921 & -3.88 & -1.27 &  0.57 & 4.93 &  8.640 & K2.3 & 0.000831 &  0.050 & 1.036 & 8.00 & 1.24  & --  & Y \\
			6519589007610428672 &  44624744 & 341.3107283 & -45.5463621 &  8.797 &  8.376 & 0.694 & 1.183 & 3.242 & 12.0772 & 21.4619 & -5.46 & -6.30 &  0.14 & 5.78 & 10.523 & F8.0 & 0.009635 & -0.050 & 0.915 & 2.95 & 0.98  & --  & Y \\
			6529714990008850176 & 382427867 & 350.4638171 & -44.2037163 &  9.128 &  8.719 & 0.671 & 1.183 & 1.016 &  7.2204 & 27.0466 & -3.11 & -2.81 &  0.18 & 1.64 &  8.059 & F7.0 & 0.011227 & -0.020 & 1.018 & 3.03 & 0.08  & --  & Y \\
			2332274963803275392 & 304698914 & 353.0001356 & -26.3089483 & 15.162 & 13.949 & 2.867 & 1.506 & 0.827 & 10.9470 & 31.3109 & -2.15 & -2.35 & 12.29 & 1.17 & 13.438 & M3.8 & 0.001062 &    nan & 1.127 & 1.94 & 0.13  & --  & Y \\
			6615156359271326464 & 441157622 & 332.9597596 & -29.5284305 &  8.972 &  8.561 & 0.675 & 1.184 & 2.787 & 18.4267 & 31.9268 & -8.52 & -0.20 &  0.16 & 3.08 & 10.423 & F7.2 & 0.015755 & -0.070 & 0.907 & 2.33 & 0.25  & --  & Y \\
			6588291854231350784 & 204590871 & 328.5881079 &  -35.061741 & 13.826 & 12.787 & 2.202 & 1.382 & 2.313 & 19.8257 & 34.3682 & -9.45 & -0.67 &  1.91 & 2.46 & 10.840 & M1.7 &      nan &  0.220 & 1.076 & 3.95 & 0.14  & --  & Y \\
			2399758825692435072 & 204291036 & 344.8686708 & -18.8613737 & 11.071 & 10.398 & 1.218 & 1.229 & 4.307 & 19.7428 & 35.6149 & -4.82 & -3.07 &  0.34 & 4.58 & 11.785 & K3.1 & 0.000456 &  0.000 & 0.869 & 8.28 & 1.33  & --  & Y \\
			2401820341275192576 &  69925665 &  339.736044 & -18.9842114 &  9.848 &  9.289 & 0.975 & 1.201 & 4.730 & 21.6717 & 40.1599 & -6.58 & -1.25 &  0.23 & 4.74 & 12.661 & G9.8 & 0.001384 & -0.030 & 1.076 & 5.55 & 0.50  & --  & Y \\
			6812532577791480448 & 270306169 & 331.2534856 & -24.1746367 & 14.706 & 13.574 & 2.547 & 1.446 & 0.487 & 22.7678 & 40.3475 & -9.24 & -2.91 &  4.52 & 1.47 & 12.029 & M3.2 &      nan &  0.360 & 1.124 & 2.40 & 0.44  & --  & Y \\
			... & ... & ... & ... & ... & ... & ... & ... & ... & ... & ... & ...&... & ... & ... & ... & ... & ... & ... &... & ... & ... & ... & ... \\
			\enddata
			\tablecomments{1}{Column 1-20 are Gaia DR3 ID, TESS Input Catalog ID (TIC), RA and DEC, $G$ magnitude,$R_p$, $B_p-R_p$, $B_p$/$R_p$ flux excess, tangential velocity offset ($V_{off}$, km s$^{-1}$), spatial separation (degree), 3D distance (pc), predicted RV (Vr(pred), km s$^{-1}$), observed RV (Vr(obs), km s$^{-1}$), error in RV (Vrerr, km s$^{-1}$), total velocity offset ($V_{tot}$), parallax (plx, mas), spectral type (SpT), F$_{NUV} $/F$_J$, $W1-W3$, and RUWE, respectively. Columns 21-22 are rotational period and errors derived by using their stellar light curves, column 23 comments on whether light curve is available for the given star and whether short-period rotational signal is detected. Column 24 judges whether the star is a sibling of TOI-251, ``Y" means sibling, ``N" means non-sibling, ``U" means uncertain sibling.}
			\tablecomments{2}{Information for the 11 stellar siblings are shown here for reference. Other uncertain and non- members are included in the entire table available online.}
			\tablecomments{3}{The candidates that are rejected based on $P_{ROT}$ or Li -age are identified in Column 24 by using a superscript of ``Li" or ``P" right to ``N".}
			\tablecomments{}{(This table is available in its entirety in machine-readable form.)}
		\end{splitdeluxetable*}
		
		\subsection{A comparison of TOI-251 sibling candidates to multiple open clusters in the color-magnitude diagram (CMD)} \label{sec:CMD}
		
\begin{figure*}[ht!]
			\centering
			\includegraphics[width=1.0\textwidth]{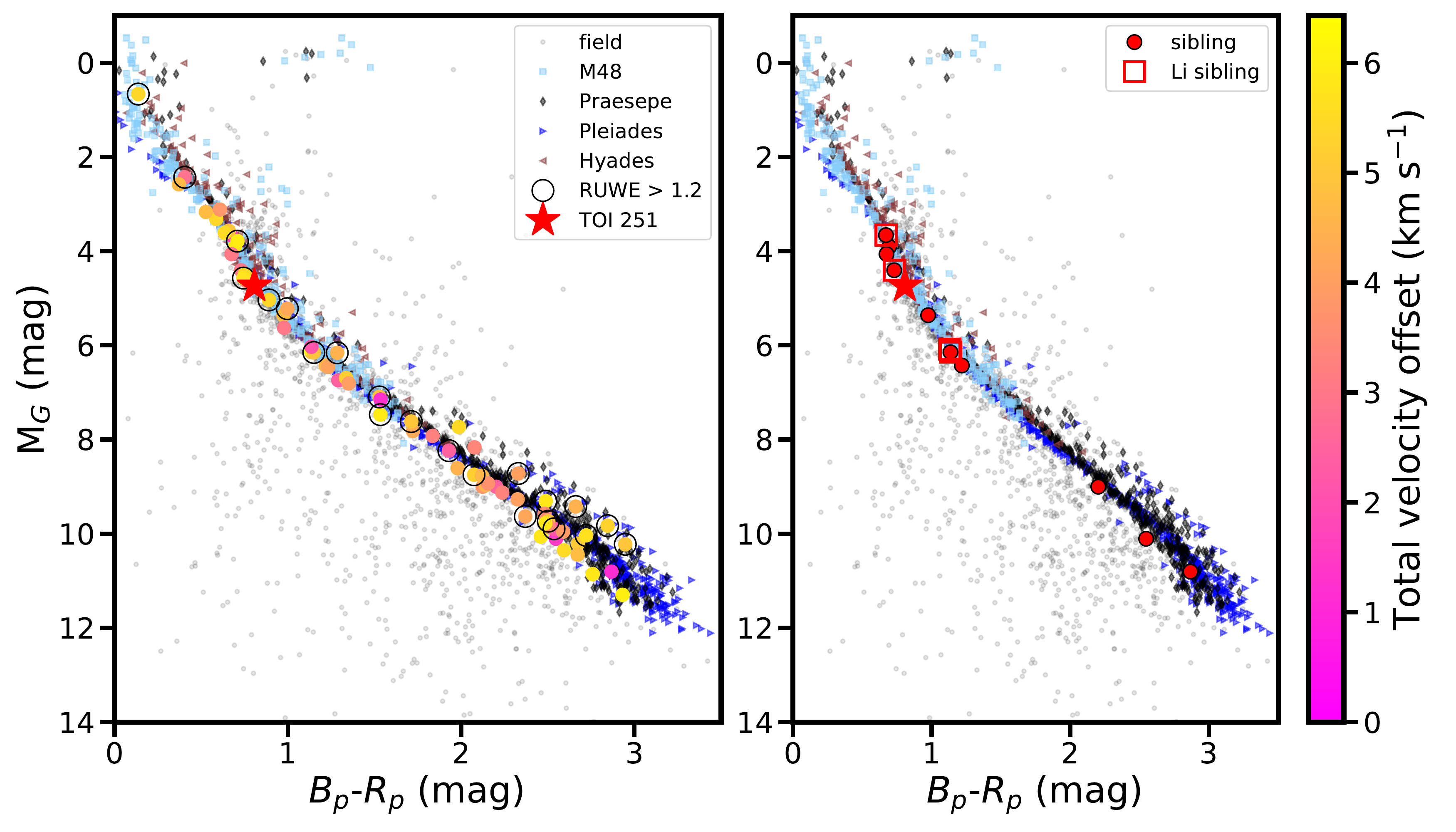}
			\caption{Color-magnitude diagram for TOI-251 siblings and comparisons to a few known clusters/associations. The TOI-251 siblings are marked in filled circles colored by their total velocity offsets ($V_{tot}$s), a $B_p/R_p$ flux excess cut of $<$ 1.6 is applied to clean up some of the highly contaminated objects at the fainter magnitude. Sibling candidates with RUWE $>$ 1.2 are likely binaries and are encircled. The Pleiades ([Fe/H] = +0.03 dex, 100--130 Myr, \citealt{2018A&A...618A..93C}), M48 ([Fe/H] = -0.06 dex, 420 Myr, \citealt{2020AJ....159..220S}), Hyades ([Fe/H] = +0.15 dex, 650 Myr, \citealt{1998A&A...331...81P}), and Praesepe ([Fe/H] = +0.15 dex, 650 Myr, \citealt{2018A&A...618A..93C}) open cluster members are shown in different colored symbols as denoted in the legend. The small black dots in the background are nearby field stars from the Gaia DR3 catalog. The right panel shows the locations of the 11 stellar siblings of TOI-251.}
			\label{fig:CMD_comp}
		\end{figure*}

		In the left panel of Figure \ref{fig:CMD_comp}, we show all TOI-251 sibling candidates in a CMD colored by their total velocity offset ($V_{tot}$, computed from RV and $V_{off}$). The right panel shows the final 11 stellar siblings of TOI-251 (discussed in Section \ref{sec:age}). We compare TOI-251 RV comoving candidates and the 11 siblings to open clusters of different age and metallicity. The Pleiades open cluster is well-studied with an age $\sim$100--130 Myr (e.g. 125 Myr, \citealt{2018A&A...613A..63B}; 130 $\pm$ 20 Myr, \citealt{2004ApJ...614..386B}; 100 Myr, \citealt{1993AAS...98..477M}); 112 $\pm$ 5 Myr, \citet{2015ApJ...813..108D}) and a metallicity very close to solar ([Fe/H] = +0.03 dex, \citealt{2021ApJ...908..119M}). The Hyades and Praesepe cluster have very similar age and metallicity ([Fe/H] = +0.15 dex, 650 Myr, \citealt{2017AJ....153..128C}). M48 has a slightly sub-solar metallicity and an age in between the Pleiades and Hyades/Praesepe ([Fe/H] = -0.063 dex, 420 Myr, \citealt{2020AJ....159..220S}). The Hyades members are shown in brown leftward triangles using data from \citealt{1998A&A...331...81P}). The Pleiades and Praesepe members are from \citet{2018A&A...618A..93C}, where we adopt a membership probability cut of $>$ 0.9. The M48 members are from \citet{2020AJ....159..220S} and \citet{2023arXiv230309783S}, shown in light sky blue squares. We apply a $B_p/R_p$ flux excess cut of $<$ 1.6 for all the clusters to clean up some of the highly contaminated objects at the fainter end.
		
		Since TOI-251 has a sub-solar metallicity of -0.106 dex (\citealt{2021AJ....161..171T}), we expect that the comoving candidates have similar sub-solar metallicity. Consistent with our expectations, the sequence of TOI-251 comoving candidates lies closer to M48, potentially because of the lower metallicity of M48 (-0.06 dex). Toward the fainter end of the CMD, the photometric scatter becomes larger, which are likely caused by field contamination and/or large photometric errors. The pinker candidates (smaller $V_{tot}$) seem to consistently fall on the bluer side of Praesepe/Pleiades, likely because the candidates have lower metallicity. We do not perform CMD fitting here because we lack subgiant/turnoff stars that are typically used for CMD fitting, and because our candidate members probably contain more contaminated field stars than typical groups of stars used for young association age dating, where the population have often been cleaned up using full 3-D kinamtics, or where the population is so numerous that contamination is less of a problem. 
		
		Based solely on the CMD, it is hard to distinguish between the young Pleiades age and the Hyades/Praesepe age of TOI-251 sibling candidates. In the next section, we perform thorough age analysis by using the gyrochronology relations, the color-lithium sequence, and the excess noise in Gaia photometry.
		
		\section{Determination of Age}  \label{sec:age}
		
		Age is one of the key stellar parameters to understanding the properties of a planetary system. The age can be determined much better if the planetary system is in a clustering environment (\citealt{2021PhDT........18S}). Stellar rotational period ($P_{ROT}$) can be derived directly from the stellar light curve (\citealt{2022RAA....22g5008S}), and one can subsequently use gyrochronology color-rotation period relations to compute age directly. Some stellar associations that have been well-studied in the literature (e.g., Pleiades, \citealt{2018A&A...613A..63B}; Hyades, Praesepe, \citealt{2017AJ....153..128C}) have well-measured stellar parameters (e.g. photometry, $P_{ROT}$), and thus one can directly compare to their $P_{ROT}$ patterns. Stellar chemical abundance, especially lithium abundance, provide an additional check for age (\citealt{2000ASPC..198..235D}). Stars destroy lithium throughout their lives and show different abundance pattern at different age. We thus could directly compare to the Lithium abundances of well-studied open clusters and young associations (e.g., \citealt{2014AJ....147..146K}, \citealt{2018A&A...613A..63B}, \citealt{2017AJ....153..128C}). In this section, we estimate age for TOI-251 sibling candidates (listed in Table \ref{tab:251friends}) using the $P_{ROT}$ -- $T_{eff}$ diagram and the gyrochronology color -- $P_{ROT}$ relation. Further, we check the Li age of selected candidates by using the color-lithium sequence; and compute an additional age estimate based on the excess noise in Gaia photometry.
		
		\subsection{Stellar Rotational Period}  \label{sec:rotp}
		
		After a star ends its pre-main sequence contraction, gyrochronology relations can be applied to compute its age using the stellar rotation period ($P_{ROT}$) and the photometric color (\citealt{2009IAUS..258..345B}). In this section, we derive $P_{ROT}$ for the TOI-251 siblings by using their light curves, compare the $P_{ROT}$ distribution with that of a few open clusters, compute the ages for individual candidates, and estimate the age for the entire group.
		
\begin{figure*}[ht!]
			\centering
			\includegraphics[width=0.8\textwidth]{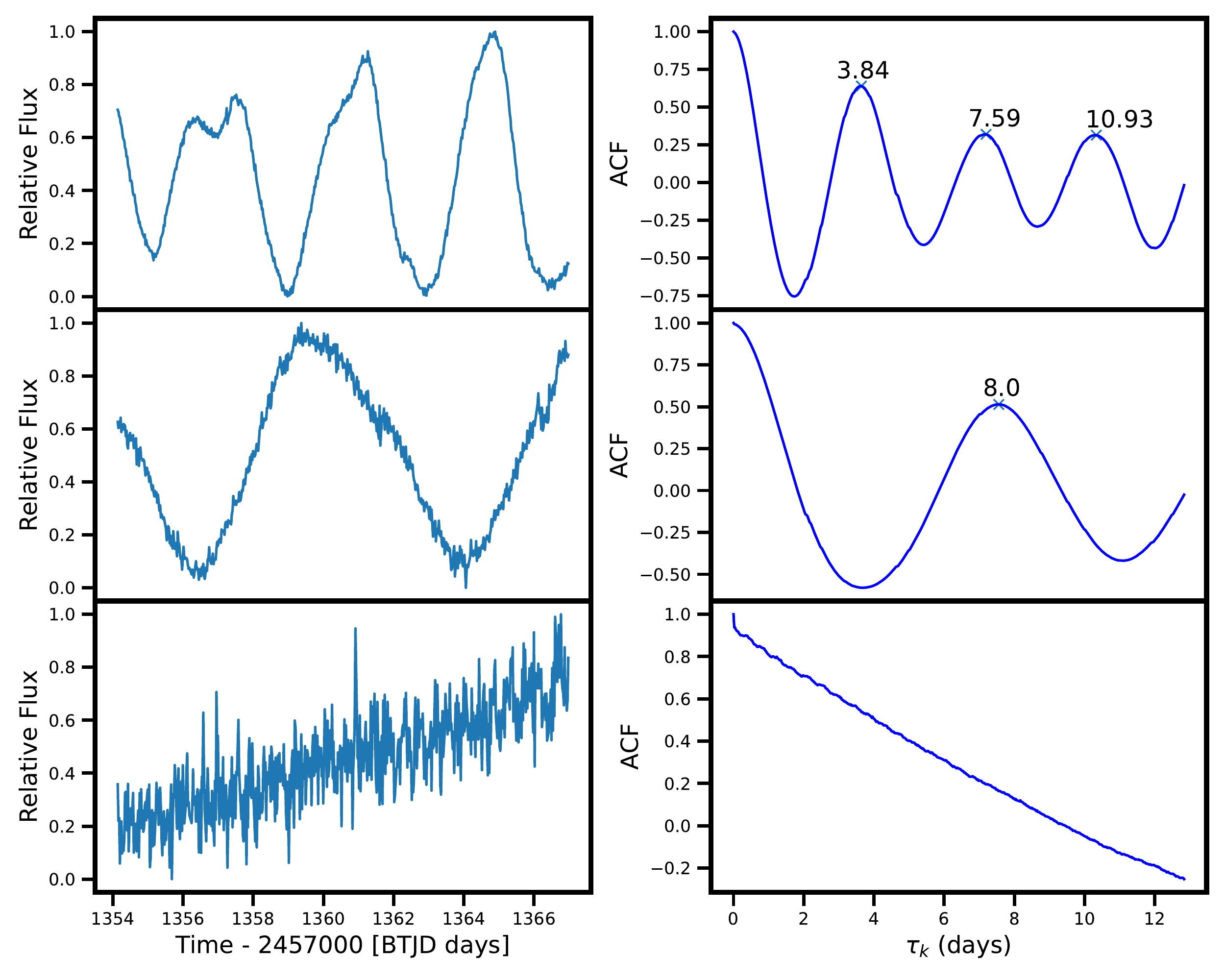}
			\caption{Light curves and autocorrelation functions for TOI-251 (or TIC 224225541; top panel) and TIC 270585410 (middle panel) as examples for how we measure stellar $P_{ROT}$ from light curves. All the peaks in the ACFs, which include the true $P_{ROT}$ and aliases, are marked in the right panel. For the two stars shown, their $P_{ROT}$  correspond to the first peak in their ACFs. The bottom panel is TIC 66473708, shown as an example of non-detection of $P_{ROT}$ .}
			\label{fig:lk_ACF}
		\end{figure*}

		We query stellar light curves for all TOI-251 sibling candidates (321 in total, including those candidates without Gaia DR3 RV) through the Mikulski Archive for Space Telescopes (MAST) portal\footnote{\url{https://mast.stsci.edu/portal/Mashup/Clients/Mast/Portal.html)}}. We use the \texttt{LightKurve} package in Python to process the raw light curves, including detrending, removing outliers, and normalizing the flux. All light curves available from MAST have been used, including the TESS Science Processing Operations Center (SPOC, \citealt{2016SPIE.9913E..3EJ}) pipeline (TESS-SPOC, \citealt{TESS-SPOC}), the TESS Asteroseismic Science Operations Center (TASOC, \citealt{2017EPJWC.16001005L, TASOC}), the QLP faint star search (\citealt{2022ApJS..259...33K, QLP}), the Kepler/K2 light curves (\citealt{10.1088/2514-3433/ab9823, Kepler, K2}), and the GSFC-ELEANOR-LITE (\citealt{2022RNAAS...6..111P, GSFC}) light curve. We calculate the autocorrelation function (ACF, \citealt{2014ApJS..211...24M}) of the light curves when they are available for each sibling candidate to estimate the stellar rotation period. Figure \ref{fig:lk_ACF} shows a few examples of light curves (left panel) and their ACFs (right panel). If a star shows a clear detection of $P_{ROT}$ signal ($<$ 12 days) in its ACF, we compute the corresponding period and error in column 21 and 22 of Table \ref{tab:251friends}. If a short $P_{ROT}$ is not detected, we give ``99.99" for the star in column 21 and 22, and give ``non-detection" in column 23; if no stellar light curve is available for this star, we also have ``99.99" for this star in both columns and in the meantime give``no lc" in column 23. Many of the candidate stars do not have detected $P_{ROT}$, which are likely field stars of older ages or that their stellar light curves are bad (e.g. low signal-to-noise ratio).
		
		Since we search a large radius (50 PC) around TOI 251, we expect much field contamination. The $P_{ROT}$ detection fraction (number of stars with detectable $P_{ROT}$/number of stars with stellar light curves available) in our field is 45.5\%, and the fraction of stars that have light curves is 69.2\%. If we restrict our calculations to the 73 comoving candidates, then the detection fraction becomes 68.1\% and the fraction of stars having light curves becomes 94.5\%. Our detection rate is much higher than that in a random field: only 14\%–16\% stars in the Kepler field have $P_{ROT}$ $<$ 18 days (\citealt{2014ApJS..211...24M, 2021ApJS..255...17S}). Since Kepler is generally expected to have better sensitivity and much longer duration of observation than TESS, the detection fraction from Kepler is overestimated when compared to that from TESS (P $<$ 12 days), making our detection fraction even more statistically robust over that in the field. We estimate the $P_{ROT}$ of TOI-251 (TIC 224225541) to be 3.84 days, consistent with the reported value of 3.84 days from its discovery paper (\citealt{2021AJ....161....2Z}). For candidates bluer than $B_p-R_p$ of about 0.5 mag, their $P_{ROT}$ measurements are probably not real (likely pulsations). We still report $P_{ROT}$ for stars bluer than 0.5 mag, and add the comment ``likely pulsation" in column 23 of Table \ref{tab:251friends}.
		
\begin{figure*}[ht!]
			\centering
			\includegraphics[width=0.9\textwidth]{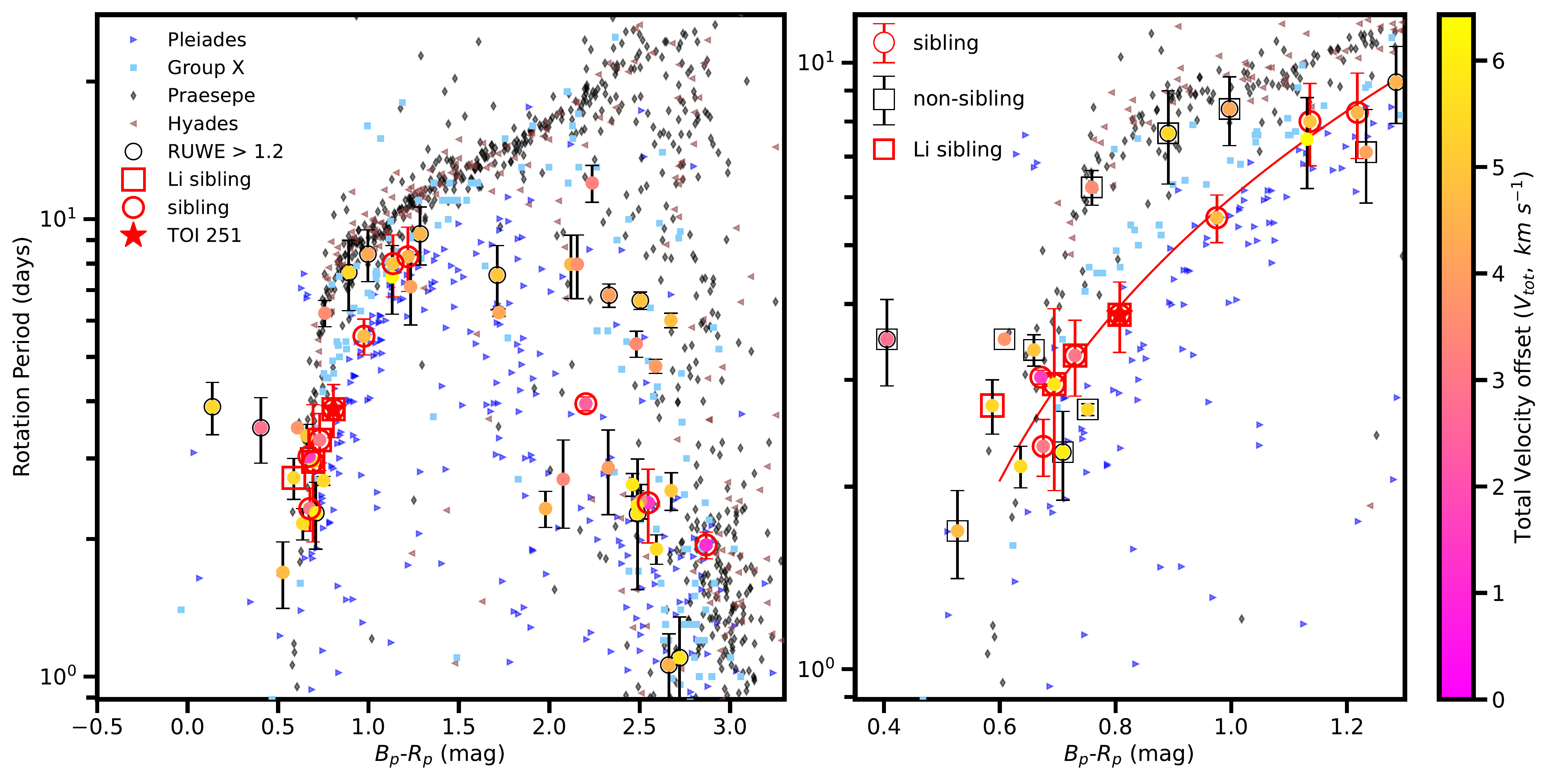} 
			\caption{Color-rotational period distribution of TOI-251 sibling candidates, plotted in filled circles colored by $V_{tot}$. TOI-251 is marked as a red star. Candidates with Gaia RUWE $>$ 1.2 are encircled. The Pleiades (\citealt{2016AJ....152..113R}), Praesepe, Hyades (\citealt{2019ApJ...879..100D, 2017AJ....153..128C}) clusters, and Group X members ($\sim$ 300 Myr, \citealt{2022AJ....164..115N}) are shown in different colored symbols as denoted in the legend. The right panel specifically zooms in the area of interest for gyrochronology with 0.35 $<$ $B_p-R_p$ $<$ 1.3 and Prot $<$ 12 days. The 11 siblings are in large red open circles, the Li-siblings are in red open squares, the red curve (right panel) fits to the siblings and clearly lie in between the Group X and Pleiades members. The non-siblings are marked in black open squares in the right panel.}
			\label{fig:TOI251_ROTP}
		\end{figure*}

		Figure \ref{fig:TOI251_ROTP} shows the color-$P_{ROT}$ distribution for TOI-251 RV comoving candidates where $P_{ROT}$ are detected (stars bluer than $B_p-R_p$ of 0.5 mag are not shown). Similarly, the Pleiades, Hyades, and Praesepe cluster are shown for comparison. We add the $P_{ROT}$ distribution of Group X members ($\sim$ 300 Myr, \citealt{2022AJ....164..115N}). The left panel shows all RV comoving candidates where $P_{ROT}$ are derived. The right panel specifically zooms in the region where gyrochronology relations apply (0.4 $<$ $B_P-R_P$ $<$ 1.3 mag). The $P_{ROT}$ of TOI-251 siblings candidates show a sequence that clearly lie in between the Pleiades and Group X (see the red line in the right panel), suggesting a homogenous age $\sim$ 200 Myr.
		
		Next, we compute age for 27 TOI-251 comoving candidates where the gyrochronology relation applies (0.4 $<$ $B_p-R_p$ $<$ 1.3 and $P_{ROT}$ $<$ 12 days). Multiple gyrochronology color-rotational period relations are available from the literature: \citet{2007ApJ...669.1167B} initially proposed the $B-V$–period gyrochronology relation, which was later calibrated by multiple groups (\citealt{2008ApJ...687.1264M, 2009ApJ...695..679M, 2011ApJ...733..115M, 2015MNRAS.450.1787A}). \citet{2019AJ....158..173A} point out that the initial \citet{2007ApJ...669.1167B} $B-V$-period relationship systematically over-predicts the ages of stars with masses $<$ 1M$_{\odot}$ and under-predicts the ages of more massive stars, which has inspired them to develop the Gaia $(B_p-R_p)$--period relation. However, this $(B_p-R_p)$--period relation, as well as some of the later ones (e.g., \citealt{2020AA...636A..76S}), have not been calibrated well for younger stars. 
		
		We retrieve $B$, $V$, $E(B-V)$, and their associated errors from the TESS Input Catalog (TIC) to calculate the de-reddened $B-V_0$ and error (from error propagation), the values are shown in column 7--8 of Table \ref{tab:rotp}. The TIC ID, $G$ magnitude, $B_p-R_p$, $V_{off}$, RV, and RUWE are shown in columns 1 -- 6 for reference. Columns 9--10 show $P_{ROT}$ and error of the star.
		
		We adopt the $(B-V)_0$--period relation from \citet{2007ApJ...669.1167B} to compute ages for the 27 stars using the calibrations from \citet{2008ApJ...687.1264M}, and from \citet{2009ApJ...695..679M}, respectively. \citet{2008ApJ...687.1264M} use younger benchmark stars from $\alpha$ Per ($\sim$85 Myr) and the Pleiades ($\sim$ 100 - 130 Myr) to calibrate the model, so is better for younger stars. We compute age and error (propagated from $P_{ROT}$ and $B-V_0$) for stars in the 0.5 $\leq$ $(B − V )_0$ $\leq$ 0.9 mag color range where the calibration applies, and add to columns 11--12 of Table \ref{tab:rotp}. Many of our stars fall out of this $(B-V)_0$ range, for which we give ``NaN". The calibration from \citet{2009ApJ...695..679M} applies to 0.5 $\leq$ (B-V)$_0$ $\leq$ 1.2 mag and $P_{ROT}$ $>$ 1.5 days, similarly, we use this calibration to compute age and error (shown in columns 13-14). 
		
		The color-age relation from \citet{2008ApJ...687.1264M} applies to nine among the 27 objects. The calibration from \citet{2009ApJ...695..679M} applies to a slightly broader $B-V_0$ color range, which adds age estimation for four more stars. Individual ages computed from the two different calibrations are consistent with each other. TOI-251 (TIC 224225541) has an age of 159/145 Myr, consistent with the observed pattern in Figure \ref{fig:TOI251_ROTP}, where it lies in between the Pleiades and Group X). The other gyrochronology age of sibling candidates also agree with their locations on the color--$P_{ROT}$ plot. In the 0.35 $<$ $B_P-R_P$ $<$ 1.3 mag range (right panel), a sequence of stars along the red curve is situated between the Pleiades and Group X, consistent with being candidates of $\sim$ 200 Myr age. We identify eight stars as TOI-251 siblings (shown as red open circles) located along the red curve with relatively small $V_{tot}$. Additionally, we classify ten stars that fall off the red curve as non-siblings (shown as black open squares), while the remaining nine are classified as uncertain siblings (five of which have no reported $P_{ROT}$). Their $P_{ROT}$ siblingship designations are shown in the last column of Table \ref{tab:rotp}.
		
		Since gyrochronology relations only apply to a limited color range, we are unable to quantitatively derive age for most of the redder ($B_P-R_P$ $>$ 1.3 mag) comoving candidates. The color-$P_{ROT}$ pattern also shows more scatters in this color range. We are thus unable to determine the siblingships for most of the stars in this region, so they are mostly identified as uncertain siblings. We identify the three pinkest stars as siblings (encircled in red in the left panel) since their $V_{tot}$ is very small and lie in between the Pleiades and Group X.
		
		In total we identify 11 TOI-251 siblings and mark them in large red open circles in Figure \ref{fig:TOI251_ROTP} (for both subplots). Though the locations of TIC 160070568 and TIC 270536278 on the $P_{ROT}$ plot agree with the red line, they have quite large $V_{tot}$ offset, and are thus assigned as ``U" (uncertain sibling). TIC 616064178 is an uncertain sibling because it has RUWE $>$ 1.2 (likely binary). TIC 12881069 does not have reported $P_{ROT}$ and is currently assigned as ``U", but later analysis of its Li-age suggests non-siblingship (marked as ``N" in Table \ref{tab:251friends}).
		
		Using TOI-251 siblings only, we compute the average age to be 204 $\pm$ 45 Myr (using the  \citet{2008ApJ...687.1264M} calibration, error based on standard deviation), or 187 $\pm$ 83 Myr (using the \citealt{2009ApJ...695..679M} calibration). Our $P_{ROT}$ and gyrochronology age analyses suggest a stellar association candidate in the Pheonix-Grus constellation because 1) the siblings' $P_{ROT}$ pattern in Figure \ref{fig:TOI251_ROTP} suggests a homogeneous age that lie in between that of the Pleiades and Group X; 2) The average age computed from gyrochronology relations supports the observed pattern; 3) and the $P_{ROT}$ detection fraction is significantly higher than that in the field.
		
		\begin{deluxetable*}{|ccccccccccccccc|}
			\tablecaption{Siblingship and gyrochronology age for TOI-251 comoving candidates} \label{tab:rotp}
			\scriptsize
			\tablehead{
				\colhead{TIC} &  \colhead{$G$} & \colhead{$B_p-R_p$}	& \colhead{$V_{off}$}	& \colhead{RV} & \colhead{RUWE} & \colhead{$(B-V)_0^1$} & \colhead{$\sigma^1$} & \colhead{$P_{ROT}$} & \colhead{$\sigma$} &\colhead{age$^2$} &\colhead{$\sigma^2$} & \colhead{age$^2$} & \colhead{$\sigma^2$} & \colhead{siblingship$^3$} \\
				\colhead{Id} & \colhead{mag} & \colhead{mag}	& \colhead{km s$^{-1}$}	& \colhead{km s$^{-1}$} &\colhead{} &\colhead{mag} &\colhead{mag} & \colhead{day} & \colhead{day} & \colhead{Myr} & \colhead{Myr}& \colhead{Myr} & \colhead{Myr} & \colhead{}
			}
			\decimalcolnumbers
			\startdata
			\multicolumn{15}{|c|}{All comoving candidates in 0.4 $<$ $B_P-R_P$ $<$ 1.3} \\
			\hline
			224225541$^4$ &  9.761 & 0.807 &   0.0 & -1.52 & 0.959 & 0.641 & 0.068 &  3.84 &  0.51 &   159 &     56 &    145 &     72 & Y \\
			175490413 &  9.638 &  0.73 & 1.425 &  1.16 & 0.948 & 0.534 & 0.053 &  3.29 &  0.47 &   258 &    212 &    314 &    298 & Y \\
			270585410 & 11.463 & 1.136 & 4.926 & -1.27 & 1.036 & 1.054 & 0.295 &   8.0 &  1.24 &   NaN &    NaN &    160 &     98 & Y \\
			44624744 &  8.797 & 0.694 & 3.242 &  -6.3 & 0.915 & 0.531 &  0.03 &  2.95 &  0.98 &   223 &    169 &    268 &    224 & Y \\
			44624757 &  8.676 & 0.709 & 4.751 & -5.14 & 3.962 & 0.554 &  0.03 &  2.28 &  0.38 &   106 &     44 &    115 &     58 & N \\
			12434574 &  8.161 & 0.587 &  3.14 & -5.93 &  0.94 & 0.453 & 0.042 &  2.72 &  0.28 &   NaN &    NaN &    NaN &    NaN & U \\
			160070568 & 11.481 & 1.131 & 4.986 &  2.54 & 0.786 & 0.947 & 0.348 &  7.48 &  1.28 &   NaN &    NaN &    174 &    147 & U \\
			12881069 $^5$ & 10.748 & 0.978 & 2.735 &  -0.2 & 0.917 & 0.843 & 0.197 &   NaN &   NaN &   NaN &    NaN &    NaN &    NaN & U \\
			382427867 &  9.128 & 0.671 & 1.016 & -2.81 & 1.018 & 0.481 & 0.102 &  3.03 &  0.08 &   NaN &    NaN &    NaN &    NaN & Y \\
			229088708 & 12.163 & 1.291 & 2.425 & -1.51 & 1.081 & 1.017 & 0.059 &   NaN &   NaN &   NaN &    NaN &    NaN &    NaN & U \\
			441157622 &  8.972 & 0.675 & 2.787 &  -0.2 & 0.907 & 0.462 &  0.04 &  2.33 &  0.25 &   NaN &    NaN &    NaN &    NaN & Y \\
			143947986 &   11.3 & 1.233 & 3.887 & -2.84 & 1.189 & 1.339 & 0.314 &  7.12 &  1.25 &   NaN &    NaN &     83 &     42 & N \\
			152864226 &  8.058 & 0.405 & 2.628 & -0.38 & 6.093 & 0.274 & 0.042 &    3.50$^6$ &  0.57 &   NaN &    NaN &    NaN &    NaN & N \\
			204291036 & 11.071 & 1.218 & 4.307 & -3.07 & 0.869 & 1.296 & 0.241 &  8.28 &  1.33 &   NaN &    NaN &    118 &     51 & Y \\
			89019293 & 11.634 &  1.15 & 4.935 & -1.49 & 1.807 & 0.904 & 0.059 &   NaN &   NaN &   NaN &    NaN &    NaN &    NaN & U \\
			616064178 & 11.271 & 1.285 & 3.401 &  1.43 & 1.217 &   NaN &   NaN &  9.29 &  1.35 &   NaN &    NaN &    NaN &    NaN & U \\
			69925665 &  9.848 & 0.975 &  4.73 & -1.25 & 1.076 &  0.87 & 0.074 &  5.55 &   0.5 &   177 &     34 &    118 &     31 & Y \\
			139148181 & 10.319 & 0.759 & 2.757 &   0.7 & 0.829 & 0.635 &   0.1 &  6.23 &  0.41 &   383 &    163 &    383 &    254 & N \\
			183984122 &  8.851 & 0.608 & 3.736 & -0.77 & 0.973 & 0.401 & 0.044 &   3.5 &   NaN &   NaN &    NaN &    NaN &    NaN & N \\
			209318498 & 11.504 & 1.136 & 0.931 & -3.45 & 0.927 & 0.862 & 0.075 &   NaN &   NaN &   NaN &    NaN &    NaN &    NaN & U \\
			382413048 &  10.75 & 0.997 & 2.244 &  2.06 & 8.952 & 0.602 & 0.121 &  8.39 &  1.09 &   757 &    521 &    865 &    883 & N \\
			28001433 &  8.631 & 0.659 & 1.844 &  3.02 & 0.926 & 0.433 &  0.03 &  3.36 &   0.2 &   NaN &    NaN &    NaN &    NaN & N \\
			270536278 &  8.146 & 0.636 &  4.34 &  1.89 & 0.915 & 0.477 & 0.042 &  2.16 &  0.17 &   NaN &    NaN &    NaN &    NaN & U \\
			5774692 &  9.954 & 0.752 & 3.904 &  2.27 & 1.074 & 0.572 & 0.091 &  2.68 &  0.06 &   121 &     82 &    127 &    123 & N \\
			176255396 &  9.067 & 0.744 & 4.432 &  2.07 & 1.602 & 0.553 & 0.095 &   NaN &   NaN &   NaN &    NaN &    NaN &    NaN & U \\
			126766627 &  8.154 & 0.527 & 2.304 & -5.65 & 1.059 & 0.395 & 0.029 &  1.69 &  0.28 &   NaN &    NaN &    NaN &    NaN & N \\
			158551107 & 10.505 & 0.891 & 4.858 & -3.94 &   1.3 & 0.784 &  0.09 &  7.65 &  1.34 &   363 &    129 &    285 &    130 & N \\
			\hline
			\multicolumn{15}{|c|}{Siblings in $B_P-R_P$ $>$ 1.3$^*$} \\
			\hline
			304698914 & 15.162 & 2.867 & 0.827 & -2.35 & 1.127 & 1.803 & 0.259 & 1.94 & 0.13 & NaN & NaN & NaN & NaN & Y\\
			204590871 & 13.826 & 2.202 & 2.313 & -0.67 & 1.076 & 1.438 & 0.141 & 3.95 & 0.14 & NaN & NaN & NaN & NaN & Y\\
			270306169 & 14.706 & 2.547 & 0.487 & -2.91 & 1.124 & 1.237 & 0.235 &  2.4 & 0.44 & NaN & NaN & NaN & NaN & Y \\
			\enddata
			\tablecomments{1}{De-reddened $(B-V)_0$ computed based on $B$, $V$, and $B-V$ color excess from the TESS Input Catalog. Errors are propagated from $B$, $V$, and $B-V$.}
			\tablecomments{2}{Age in Myr based on gyrochronology relations from \citet[][Column 11]{2008ApJ...687.1264M} and \citet[][Column 13]{2009ApJ...695..679M}. ``NaN" is shown if the gyrochronology relation is not applicable. Errors are propagated from period and $(B-V)_0$ (Columns 12 and 14).}
			\tablecomments{3}{Siblingship: ``Y" means sibling; ``U" means uncertain sibling, ``N" means non-sibling.}
			\tablecomments{4}{The first line is TOI-251.}
			\tablecomments{5}{The final siblingship designation has been changed to ``N" after checking its Li age.}
			\tablecomments{6}{For candidates bluer than 0.5 mag, the reported $P_{ROT}$ are likely not real but pulsations.}
			\tablecomments{*}{All the other comoving candidates in this color range are uncertain siblings.}
		\end{deluxetable*}
		
		\subsection{The color-lithium sequence} \label{sec:li}
		
		The Lithium depletion pattern, as seen in a variety of stellar associations at different ages, suggests that lithium abundance is highly dependent on age (\citealt{2000ASPC..198..235D}). Lithium abundance (or equivalent width, EW) thus provides a direct observational tool to estimate the age of a star or an association.
		
		We obtained spectra from the Network of Robotic Echelle Spectrographs (NRES; \citealt{2018SPIE10702E..6CS}) at the Las Cumbres Observatory Global Telescope (LCOGT) network for five TOI-251 sibling candidates brighter than $G = 11$ mag, which all have $M_G$ $<$ 6 mag in Figure \ref{fig:CMD_comp}. These brighter stars have less photometric scatter, and their lithium EW pattern in this color range is more indicative of age. The NRES spectra have a spectral resolution of R$\sim$ 53000 and span 3800-–8600 \AA\ in wavelength. We processed NRES spectra by using the standard BANZAI pipeline (\citealt{2018SPIE10707E..0KM}, see description in \url{https://banzai-nres.readthedocs.io/en/latest/banzai_nres/data-products.html}), and we used the IRAF\footnote{IRAF is distributed by the National Optical Astronomy Observatories, which are operated by the Association of Universities for Research in Astronomy Inc., under cooperative agreement with the National Science Foundation.} {\it splot} tool to measure lithium EW for anyone with an NRES spectrum. In Table \ref{tab:lithium} (column 6), we show our lithium EW measurements for selected stars. Typical errors for Li EW measurement are 10-20 m\AA.
		
		\begin{deluxetable*}{cccccccccc}
			\tablecaption{Li EW and age posterior for selected candidates} \label{tab:lithium}
			\scriptsize
			\tablehead{
				\colhead{Simbad$^1$} & \colhead{TIC} & $B_p-R_p$ & $B-V_0^2$ & $P_{ROT}$ & Li EW$^3$ & age$^4$ & 1 $\sigma$ range$^4$ & 2$\sigma$ range$^4$ & Li$^5$\\
				\colhead{name} & \colhead{Id} & mag & mag & membership & m\AA\ & Myr & Myr & Myr & siblingship
			}
			\decimalcolnumbers
			\startdata
			CD-37 15287 (TOI-251) & 224225541 & 0.807  & 0.641 & Y & 134 & 210 & 101 - 504 & 23.4 - 745 & Y \\
			HD 221760 & 144276313 & 0.138 & 0.086 & -- & 6* & -- & -- & -- & U \\
			HD 798 & 12434574  & 0.587 & 0.453 & U & 60 & 294 & 141 - 441 & 41.4 - 577 & Y$^6$ \\
			HD 215355 & 44624744  & 0.694 & 0.531 & Y &139 & 220 & 53.9 - 573 & 13.1 - 762 & Y \\
			CD-38 15495 & 175490413 & 0.730 & 0.534 & Y & 131 & 230 & 64 - 577 & 16.4 - 763 & Y \\
			CD-26 16423 & 12881069 & 0.978 & 0.843 & U & 33 & 437 & 331 - 596 & 168 - 753 & N \\
			\enddata
			\tablecomments{1}{Name from the Simbad Astronomical database (\citealt{2000AAS..143....9W})}
			\tablecomments{2}{$(B-V)_0$ from the TIC catalog.}
			\tablecomments{3}{Lithium equivalent width (EW) (contribution from the Fe I 6707.4 \AA\ line included).}
			\tablecomments{4}{Median age, the 1 $\sigma$ age range, and the 2$\sigma$ age range from the BAFFLE age posterior.}
			\tablecomments{5}{Whether the star is an association member based on its Li age.}
			\tablecomments{6}{The final membership designation for this star is uncertain because its $P_{ROT}$ pattern suggests uncertain membership.}
			\tablecomments{*}{We recommend the readers to treat EW $<$ 10 m\AA\ as upper limit.}
		\end{deluxetable*}
		
\begin{figure}[ht!]
			\centering
			\includegraphics[width=0.45\textwidth]{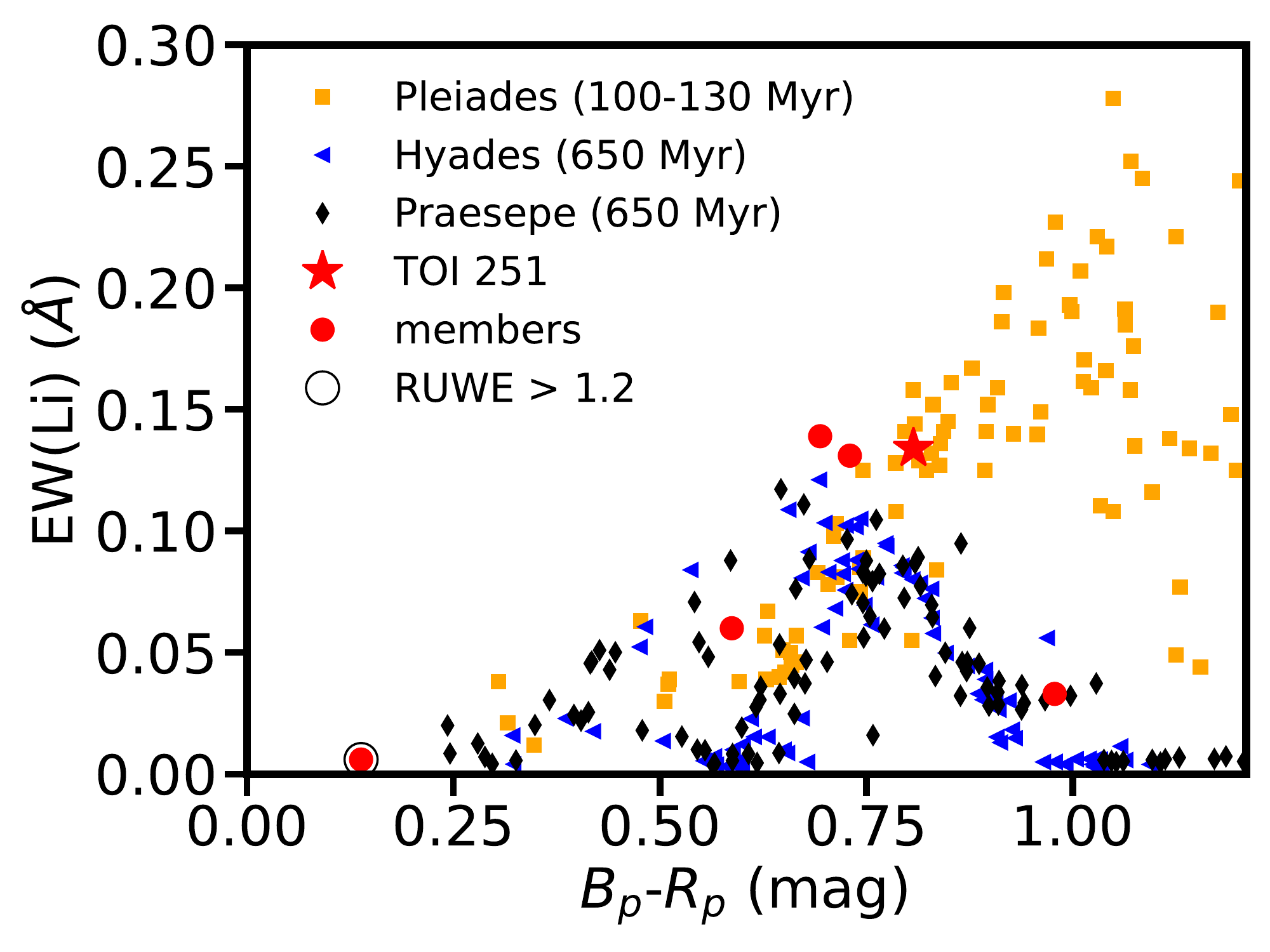} 
			\caption{Li equivalent widths (EW) of selected TOI-251 sibling candidates as a function of their Gaia $B_p − R_p$ colors. TOI-251 is shown as the red star, and Li EW for selected sibling candidates are shown in red filled circles. Candidates with Gaia RUWE $>$ 1.2 are encircled. The Pleiades (\citealt{2016AJ....152..113R}), Praesepe, and Hyades (\citealt{2019ApJ...879..100D}) open clusters are shown in different colored symbols as denoted in the plot.}
			\label{fig:lithium}
		\end{figure}

		Figure \ref{fig:lithium} shows lithium EW measured by using the NRES spectra. Lithium EWs of the Pleiades (\citealt{2016AJ....152..113R}), Praesepe, and Hyades members (\citealt{2019ApJ...879..100D}) are from the literature and plotted for comparison. TIC 144276313 (the star with the bluest $B_p − R_p$ color) has a Li EW measurement of only 6 m\AA, and we recommend that readers treat this as an upper limit. Furthermore, no cluster data lie in this color range, so we exclude TIC 144276313 from further discussions regarding the age of the association. TIC 12881069 lies closer to the Praesepe/Hyades Li sequence, so it is unlikely to be a sibling. Four stars lie apparently above the Hyades/Praesepe pattern, including TOI-251, indicating an age much younger than the Hyades/Praesepe (650 Myr) and closer to the Pleiades. The Pleiades and Hyades/Praesepe themselves have much scatters and overlap (in some regions) in between 0.25 – 0.8 mag ($B_P-R_P$), so we do not rely solely on the plot but quantitatively estimate the Li-age.
		
		We compute age using our Li EW measurement and the \texttt{BAFFLES} code. The Bayesian Age For Field LowEr mass Stars (\texttt{BAFFLES}; \citealt{2020ApJ...898...27S}) code generates age posteriors by using individual Li EW and $B-V$ color. \texttt{BAFFLES} is built up from the color–Li sequences of multiple stellar associations across a wide range of stellar ages: NGC 2264 (5.5 Myr), $\beta$ Pic (24 Myr), IC 2602 (44 Myr), $\alpha$ Per (85 Myr), the Pleiades (130 Myr), M35 (200 Myr), M34 (240 Myr), Coma Ber (600 Myr), the Hyades (700 Myr), and M67 (4 Gyr). We retrieve $(B−V)_0$ colors from the TIC catalog (shown in column 4 of Table \ref{tab:lithium}), and use \texttt{BAFFLES} to derive age posteriors for each of the individual star (except for TIC 144276313, which only has an upper limit Li EW). The median, 1$\sigma$ and 2$\sigma$ confidence intervals are given in columns 7, 8, 9 of Table \ref{tab:lithium}, respectively.
		
\begin{figure}[ht!]
			\centering
			\includegraphics[width=0.45\textwidth]{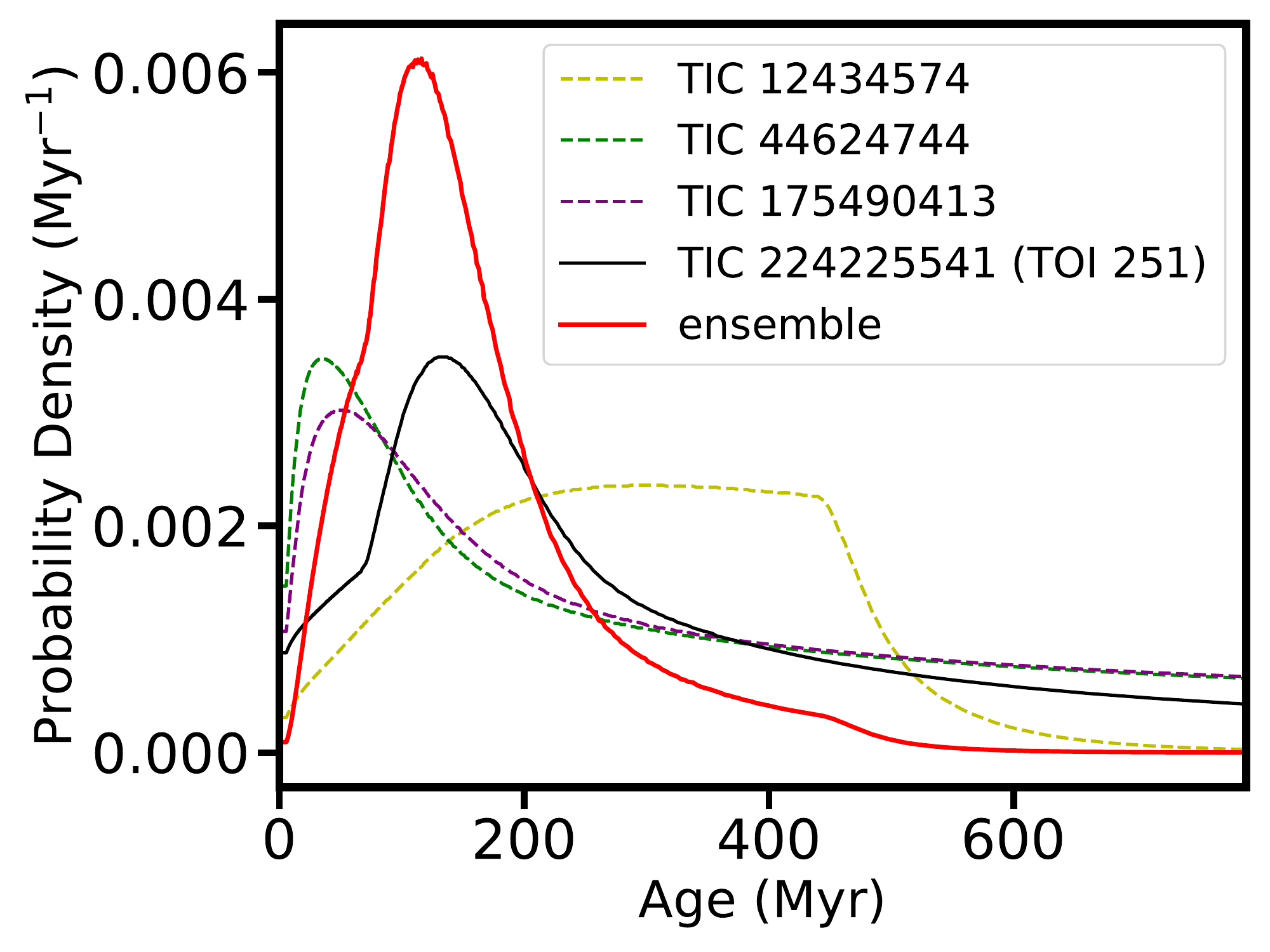} 
			\caption{Li age posteriors generated from the \texttt{BAFFLES} code. The solid black line is the age posterior for TOI 251. The other colored dash lines are age posteriors for other candidates. The solid red line is the ensemble posterior. The median age from the ensemble posterior is 238 $\pm$ 38 Myr, and the 1 $\sigma$ confidence interval is 135--420 Myr.}
			\label{fig:posterior}
		\end{figure}

		TOI-251 (TIC 224225541) has a median age of 210 Myr, slightly larger than its $P_{ROT}$ age from Section \ref{sec:rotp} (see Table \ref{tab:rotp}).  All the four stars that lie apparently above the Hyades/Praesepe members and closer to the Pleiades members have Li age that agree with the averaged $P_{ROT}$ age, so we assign the four as Li members in Table \ref{tab:lithium}, mark them in Figure \ref{fig:TOI251_ROTP} in red open squares, and mark in Column 24 of Table \ref{tab:251friends}. TIC 12434574 has uncertain membership based on its location on the color-$P_{ROT}$ diagram, so we assign ``U" as its final membership. TIC 12881069 has an older Li age than that of the association, it also lies closer to the Praesepe/Hyades Li sequence, so it is assigned as a non-sibling; the star also do not have measurable $P_{ROT}$ (``U" from Section \ref{sec:rotp}) so we assign it as a non-sibling in Table \ref{tab:251friends}. We multiply the individual posteriors together to derive an ensemble posterior of the population, TIC 12881069 has been excluded from the calculation. In Figure \ref{fig:posterior}, we show the age posterior of TOI-251 in a solid black line, age posteriors of the other sibling candidates in dash colored lines, and the ensemble posterior in a solid red line. From the ensemble posterior, the median Li-age is 238 $\pm$ 38 Myr, and the 1 $\sigma$ confidence interval is 135 -- 420 Myr, consistent with the average gyrochronology age of 204 $\pm$ 45 Myr (or 187 $\pm$ 83 Myr) within 1 $\sigma$. 
		
		Stars deplete Li throughout their lives, so in general the younger stars contain more Li in their atmosphere than older stars (\citealt{2022MNRAS.513.5387S}). In the meanwhile, Li is dependent on many other parameters, such as the metallicity and rotational velocity (\citealt{2021PhDT........18S}). Lower metallicity stars are capable of preserving more Li than similar higher metallicity stars. In Pleiades and M35, the rapid rotating cool dwarfs preserve more Li (\citealt{2018AJ....156...37A, 2021MNRAS.500.1158J}). Even with some very well defined open cluster samples, their Li pattern can show large scatter. Since the number of stars we use to estimate Li age is small, and the \texttt{BAFFLES} code do not consider more complicated stellar parameters (metallicity, rotational velocity, etc.), we use Li EW only as an additional check on (and supplementary to) the gyrochronology age.
		
\subsection{Gaia Variability Age}
\label{sec:gaia_age}
		
We compute an additional age estimate based on the excess noise in {\it Gaia} photometry. As shown in \citet{2023arXiv230209084B}, younger stars generally show higher levels of photometric variability than expected based on their magnitudes. Although this varies significantly from star to star, the 90th-percentile within a distribution is tightly correlated to age. Using the \citet{2023arXiv230209084B} calibration\footnote{\url{https://github.com/madysonb/EVA}} yielded an age of $193^{102}_{-54}$\,Myr. This is consistent with our other age estimates.
		
\subsection{A potential $\sim$200 Myr young association candidate in the Phoenix-Grus constellation} \label{sec:new_association}
		
The color -- $P_{ROT}$ sequence and the quantitative age computed from gyrochronology relations of TOI-251 stellar siblings suggest a homogenous age. Furthermore, the detection fraction of stars that have short stellar rotational period measurement in TOI-251 neighbors is 68.1\%, a statistically robust excess above that in a typical field star population (e.g. 14\% -- 16\% from Kepler, \citealt{2014ApJS..211...24M, 2021ApJS..255...17S}). The additional check on the candidates' Li age and Gaia variability age agree with that from gyrochronology. These evidences suggest a potential young association candidate in the Pheonix-Grus constellation. In the right panel of Figure \ref{fig:CMD_comp} we show the 11 siblings in red filled circles, which form a clear sequence in the CMD and suggest slightly sub-solar metallicity for the members in general.

Unfortunately, the siblings of TOI-251 have larger radial and tangential velocity offsets than typical known young associations. The previous MELANGE groups identified using this method have had the benefit of being dynamically cold and still spatially concentrated, such that the existence of a group was pretty compelling to visual inspection. The closest case is MELANGE-1, which is closer in age ($\sim$ 250 Myr, \citealt{2021AJ....161..171T}) and hence has similar large $V_{off}$ scatters. Most of the stars in MELANGE-2 (\citealt{2022AJ....164..115N}) and MELANGE-3 ((\citealt{2022AJ....164...88B}) that agree in RV space are within 1-2 km s$^{-1}$ in $V_{off}$s.
		
We consider the candidate less in terms of dispersing a Beta Pictoris moving group (\citealt{2014MNRAS.445.2169M}) or some of the tightly-bounded MELANGE groups, and more like all of the Sco-Cen association (10-20 thousand members, \citealt{2011MNRAS.416.3108R}). In 200 Myr, Sco-Cen will be quite dispersed, but the stars might be able to spread over a 500 x 500 x 100 parsec box. That means a given volume with radius 25 parsecs might still contain 25-50 members. However, the members would not represent a coherent subclump, but rather the randomly dispersed mix of everything. In that case, a larger velocity dispersion would probably be natural to expect, because it is not likely the two stars that are adjacent now were adjacent for all 200 Myr. Each are on their own orbit through the Milky Way, and they just happened to converge here at this moment. Future analysis like taking Sco-Cen and evolving it forward in the Milky Way potential would help to test this hypothesis.
		
\citet{2023arXiv230209084B} also provide a metric to judge if an association is real using the same variability information used to estimate the age. The test compares the number of level of variability for the group compared to random draws of stars at the same distance as the candidate group. Following their method, we found $\log{K}=3$, which indicates the association has more variable stars than expected from the field ($\log{K}>0.5$ is considered passing). However, this value is low compared to many of the other MELANGE and Theia associations, which often have $\log{K}>10$.
		
In Section \ref{sec:rotp} and  \ref{sec:li}, we thoroughly discuss the procedure and criteria for determing memberships for TOI-251 sibling candidates. Our final siblingship designation is shown in the last column of Table \ref{tab:251friends} (``Y" for sibling; ``N" for non-sibling; ``U" for uncertain sibling). We mark the non-siblings that are rejected based on $P_{ROT}$ or Li -age by adding a superscript to the right of ``N" (N$^{P}$ or N$^{Li}$). There are a number of uncertain siblings because we lack Gaia DR3 RV for them, or that they are located in the color range where the gyrochronology relations do not apply. Future RV measurements and/or improvements in empirical gyrochronology relations would help to determine their siblingships. In addition, if a star does not have a short $P_{ROT}$ detected in its light curve, it could be either that the star is photometrically quiet with a long rotation period, or that the light curve has too low an SNR which makes the $P_{ROT}$ hard to be detected, these stars are also assigned as ``U". Similarly, for star that does not have any light curve available, its siblingship remains uncertain (``U").
		
\begin{figure*}[ht!]
			\centering
			\includegraphics[width=1.0\textwidth]{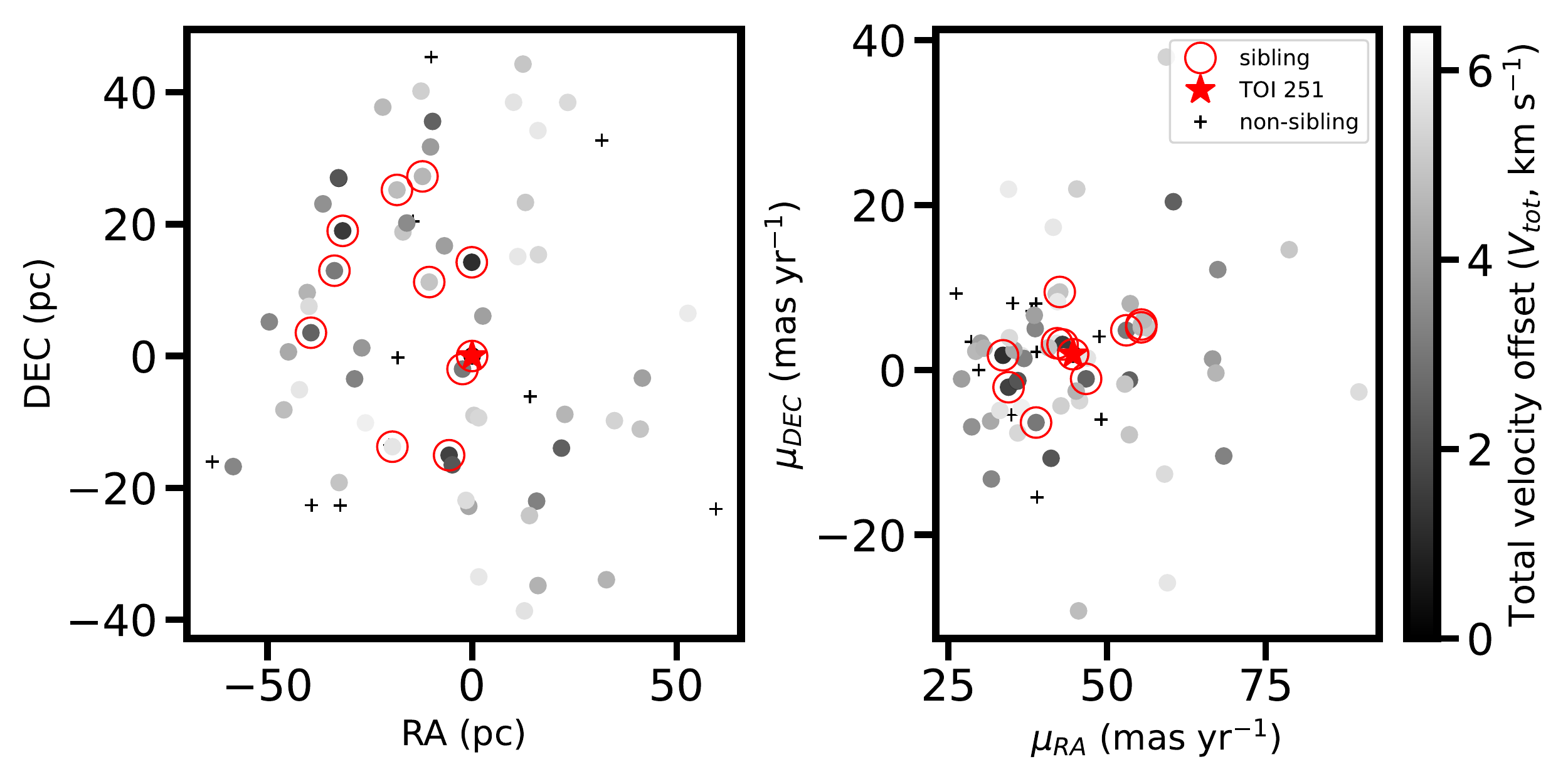} 
			\caption{The spatial (left panel) and proper motion (right panel) distributions of TOI-251 siblings (candidates). Siblings and uncertain siblings are in filled circles, colored by their $V_{tot}$s in gray scale. The 11 siblings are in red open circles, and the non-members are in plus signs.}
			\label{fig:association_dist}
		\end{figure*}

In Figure \ref{fig:association_dist}, we show the spatial and proper motion distributions of TOI-251 siblings. Siblings and uncertain siblings are in filled circles colored by their $V_{tot}$s, the non-siblings are in plus signs. The siblings are in a more gathered region, and many uncertain siblings in darker colors (small $V_{tot}$) lie close to the siblings, suggesting that many of them are likely to be siblings. Future information on the age of the uncertain candidates would help confirm their memberships.
		
\section{Summary and Discussion} \label{sec:discussion}
		
TOI-251 b is a recently discovered planet without knowingly membership to any co-moving groups, leading to an imprecise age estimation of 40 -- 320 Myr. Motivated by its potentially young age, we use the \texttt{Comove} Python package to search for TOI-251's siblings based on kinematics and spatial proximity. We use the CMD, the color--$P_{ROT}$ sequence, and the color-Li sequence to select siblings candidates and estimate the age. The 11 siblings form a clear, slightly metal-poor sequence on the CMD. Using all available light curves from MAST, we derive $P_{ROT}$ and compare to the distributions of known open clusters. The siblings show a well-defined $P_{ROT}$ sequence (0.4 $<$ $B_P-R_P$ $<$ 1.3 mag) that lie in between the Pleiades (100 Myr) and Group X (300 Myr) members, suggesting a homogenous age $\sim$ 200 Myr. The observed pattern is consistent with the average gyrochronology age of 204 $\pm$ 45 Myr (or 187 $\pm$ 83 Myr), where the age for individual stars are computed by using empirical gyrochronology relations. The $P_{ROT}$ detection fraction is a statistically robust excess above the expected field star population. In addition, the Li age and Gaia variability age are in agreement with the gyrochronology age. Our analyses also suggest a potential young association candidate in the Pheonix-Grus constellation. Similar procedures can be applied to find more stellar siblings of planetary systems and place better constraints on their age.
		
\begin{figure}[ht!]
			\centering
			\includegraphics[width=0.45\textwidth]{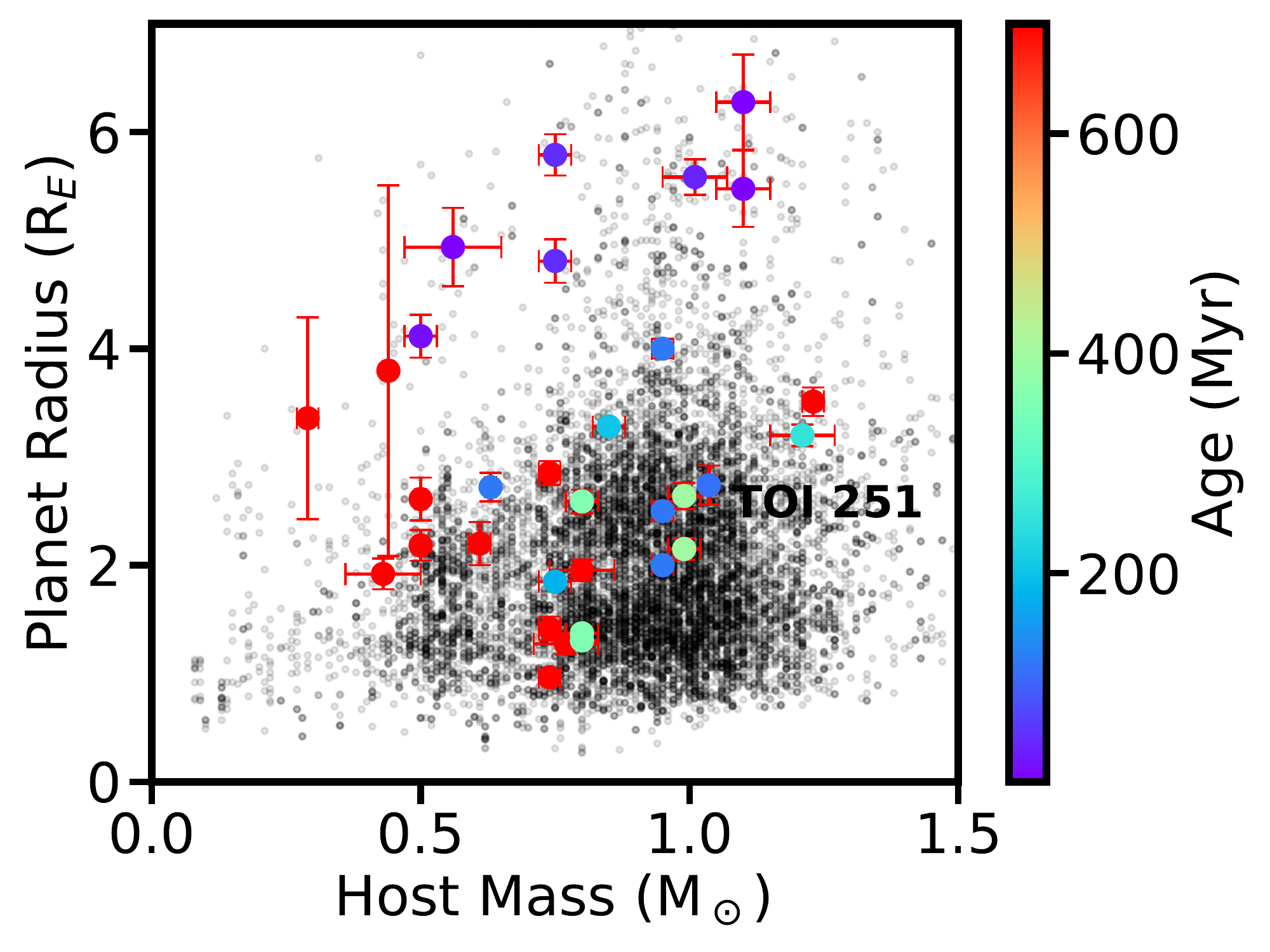} 
			\caption{Distribution of young planets and comparison to their field-age counterparts, the planet radius ($R_E$) is plotted as a function of host star mass ($M_{\odot}$). Young planets ($<$ 700 Myr) are shown in large filled circles color-coded by their ages. Field-age planets are shown in small black dots in the background, using data from the exoplanet archive (\url{exoplanetarchive.ipac.caltech.edu/}). TOI-251 is marked on this figure, which has a radius larger than most of its field-age counterparts.}
			\label{fig:rad_dist}
		\end{figure}

TOI-251b, with an age $\sim$ 200 Myr, adds to the group of young planets in stellar associations. Empirically, the radii of young planets are statistically larger than that of their field-age counterparts (\citealt{2018ApJ...855..115B, 2018AJ....155....4M, 2018AJ....156..195R}). We show planet radii as a function of host star mass in Figure \ref{fig:rad_dist} for all confirmed short-period planets with P $<$ 30 days. For field-age planets we use data from the NASA Exoplanet Archive (\url{exoplanetarchive.ipac.caltech.edu/}), which are shown as small gray circles in the background. We include a number of planetary systems discovered in the Hyades and Praesepe cluster ($\sim$ 650 Myr; \citealt{2018AJ....156..195R, 2016ApJ...818...46M, 2017AJ....153...64M, 2018AJ....155....4M}), several planets in young associations (\citealt{2016AJ....152...61M, 2020Natur.582..497P, 2019ApJ...885L..12D, 2021AJ....162...54H}), all the THYME planets (\citealt{2019ApJ...880L..17N, 2020AJ....160...33R, 2020AJ....160..179M, 2021AJ....161...65N, 2021AJ....161..171T, 2022AJ....163..156M}), three young planets orbiting K2-233 (\citealt{2018AJ....155..222D}), and several young planets in the Kepler field (\citealt{2022AJ....163..121B, 2022arXiv220501112B, 2022AJ....164...88B}). The young planets are in large filled circles colored (large circles in Figure \ref{fig:rad_dist}) by their age. And then, we add TOI-251 ($\sim$ 200 Myr) to this figure, where the host star mass and planet radius are 1.036$^{+0.013}_{0.009}$ $R_{\odot}$, and 2.74$^{+0.18}_{0.18}$ $R_E$ (\citealt{2021AJ....161....2Z}), respectively. TOI-251 b has a radius larger than most of its field-age counterparts, however, we are uncertain if TOI-251 b is inflated due to a lack of knowledge on the planet's mass.
		
		\section*{}
		
		We thank an anonymous referee for the constructive suggestions which helped us to improve this work.
		
		Q.S. thanks support from the Shuimu Tsinghua Scholar Program. We thank Luke G. Bouma for his suggestions which helped to improve this work. We also thank Hongjing Yang for his help with the TESS photometry. This work is partly supported by the National Science Foundation of China (Grant No. 12133005).
		
		This research has made use of the NASA Exoplanet Archive, which is operated by the California Institute of Technology, under contract with the National Aeronautics and Space Administration under the Exoplanet Exploration Program.
		
		This paper includes data collected by the TESS mission, which are publicly available from the Mikulski Archive for Space Telescopes (MAST). Funding for the TESS mission is provided by NASA’s Science Mission directorate. This research has made use of the Exoplanet Follow-up Observation Program website, which is operated by the California Institute of Technology, under contract with the National Aeronautics and Space Administration under the Exoplanet Exploration Program.
		
		This work has made use of data from the European Space Agency (ESA) mission Gaia, processed by the Gaia Data Processing and Analysis Consortium (DPAC). Funding for the DPAC has been provided by national institutions, in particular the institutions participating in the Gaia Multilateral Agreement.
		
		This research has made use of the VizieR catalogue access tool, CDS, Strasbourg, France. The original description of the VizieR service was published by \citet{2000AAS..143...23O}. Resources supporting this work were provided by the NASA High-End Computing (HEC) Program through the NASA Advanced Supercomputing (NAS) Division at Ames Research Center for the production of the SPOC data products.
		
		We acknowledge the use of public TOI Release data from pipelines at the TESS Science Office and at the TESS Science Processing Operations Center. We acknowledge the use of public TESS data from pipelines at the TESS Science Office and at the TESS Science Processing Operations Center.
		
		Facilities: Exoplanet Archive, TESS
		Software: matplotlib (\citealt{2007CSE.....9...90H}), BANYAN $\Sigma$ (\citealt{2018ApJ...856...23G}).
		
		\bibliography{toi251}{}
		\bibliographystyle{aasjournal}
		
		
		
	\end{document}